\documentclass[%
 reprint,
superscriptaddress,
showpacs,
 amsmath,amssymb,
 aps,
 pra,
]{revtex4-1}

\usepackage{graphicx}
\usepackage{dcolumn}
\usepackage{bm}
\usepackage{epstopdf}
\usepackage{xcolor}

\usepackage{xspace}
\newcommand{\ket}[1]{\ensuremath{|#1\rangle}\xspace}

\def \buildrelum#1\over#2{\mathrel{\mathop{\kern0pt #2}
                     \limits_{#1}   } }
\def \widetil{\mathrel\mathchar"0367}
\def \widesim{~~~~\lower.40em\hbox{$\widetil$}\mskip-10mu\succ ~~~}
\def \goes(#1){ \buildrelum #1 \over \longrightarrow}
\def \aslim(#1) {\buildrelum #1 \over \widesim}

\usepackage{dsfont}
\newcommand{\mean}[3]{\left\langle#1\middle|#2\middle|#3\right\rangle}
\newcommand{\G}{\bar{\bar{G}}}

\newcommand{\ff}{\bar{\bar{f}}}
\newcommand{\F}{\bar{\bar{F}}}

\begin{document}


%
\title{Critical review of quantum plasmonic models for finite-size media}

\affiliation{Laboratoire Interdisciplinaire Carnot de Bourgogne, CNRS UMR 6303, Universit\'e Bourgogne Franche-Comt\'e,
BP 47870, 21078 Dijon, France}
\author{V. Dorier}
\affiliation{Laboratoire Interdisciplinaire Carnot de Bourgogne, CNRS UMR 6303, Universit\'e Bourgogne Franche-Comt\'e,
BP 47870, 21078 Dijon, France}
\author{S. Gu\'erin}
\affiliation{Laboratoire Interdisciplinaire Carnot de Bourgogne, CNRS UMR 6303, Universit\'e Bourgogne Franche-Comt\'e,
BP 47870, 21078 Dijon, France}
\author{H. R. Jauslin}
\email{jauslin@u-bourgogne.fr}
\affiliation{Laboratoire Interdisciplinaire Carnot de Bourgogne, CNRS UMR 6303, Universit\'e Bourgogne Franche-Comt\'e,
BP 47870, 21078 Dijon, France}

\date{\today}

\begin{abstract}
We provide a critical analysis of some of the commonly used theoretical models to describe quantum plasmons. We summarize the standard approach based on a Fano diagonalization and we show explicit discrepancies in the obtained results by taking the limit of vanishing coupling between the electromagnetic field and the material medium. We then discuss the derivation of spontaneous emission in a plasmonic environment, which usually relies on a Green tensor and is based on an incomplete identity. The effect of the missing terms is calculated in a one-dimensional model.
\end{abstract}
 \pacs{42.50.Nn, 71.45.Gm, 42.50.Ct }
 


\maketitle
\section{Introduction}

With the development of quantum sources and single-photon detectors on one hand, and classical plasmonics on the other hand, new theories have been developed to describe the behavior of light when interacting with dissipative and dispersive media (typically, metals). One can date the first article on the matter back to 1992, by Huttner and Barnett \cite{Huttner1992}. Their goal was to introduce dissipation in the quantization of light in a homogeneous (therefore infinite) medium. It was followed by many other works \cite{HuttnerBarnett1992b,Welsch1995,Welsch1996,Welsch1998a,Welsch1998b,Welsch2000,
Welsch2007,Suttorp2004a,Suttorp2004b,Suttorp2007,Sipe2006,Philbin2010,Philbin2014,
DiStefano2001,Drezet2017a,Dorier2019a} which used two main approaches:
		\begin{itemize}
			\item a phenomenological approach formulated in terms of quantum Langevin equations first introduced by Gruner and Welsch in 1995 \cite{Welsch1995} and 1996 \cite{Welsch1996};
			\item microscopic oscillator models for the medium coupled to the electromagnetic field.
		\end{itemize}
		The latter are variations of models of the type first proposed by Hopfield \cite{Hopfield1958}, which was the approach originally used by Huttner and Barnett. Extensions of their work to inhomogeneous media were treated in \cite{Suttorp2004a,Suttorp2004b,Suttorp2007,Philbin2010}. They involve a Fano-type diagonalization in terms of bosonic creation and annihilation operators. The results of the microscopic approach were meant to provide a justification of the phenomenological quantum Langevin noise models. The main criterion for the choice of the microscopic model is that if one integrates the equations for the medium and one inserts the obtained currents into the microscopic Maxwell equations one should obtain the macroscopic Maxwell equations \cite{Sipe2006,Philbin2010}.
		
		In a recent work \cite{Dorier2019a} another quantization procedure, designed specifically to treat plasmonic systems with a finite-size medium, was formulated. Starting from the same classical model, this approach leads to results substantially different from the formulas derived and used in the past literature. In particular, it was shown to give a different spectral structure of the diagonalized system and a different expression of the electric field operator. Our interpretation of this discrepancy is that the former results \cite{Huttner1992,HuttnerBarnett1992b,Welsch1995,Welsch1996,Welsch1998a,Welsch1998b,
Suttorp2004a,Suttorp2004b,Suttorp2007,Sipe2006,Philbin2010} were obtained under the implicit hypothesis of an infinite bulk medium, although it has been extensively extrapolated to finite media in later works. Our results \cite{Dorier2019a} have some conceptual consequences and may lead to quantitative divergences from past studies. The present article aims at addressing in detail the reasons that the formulas obtained for bulk systems cannot be applied to systems with a finite medium.
		
		We first summarize the quantization procedure of Refs.~\cite{Suttorp2004a,Suttorp2004b,Suttorp2007,Philbin2010} in Section~\ref{Sec-summary}. In Section~\ref{Sec-problem} we point out explicitly how the formulas obtained in this approach lead to inconsistent results in some regimes. In Section~\ref{Sec-justification} we briefly review some criticism of the bulk approach which have been formulated in the literature \cite{DiStefano2001,Drezet2017a,Dorier2019a}. Finally in Section~\ref{Sec-green} we analyze other discrepancies that occur when calculating the decay rate of an emitter assuming the bulk formulas, which are based on Green tensors.
		
		\section{Summary of the previous approach in literature}\label{Sec-summary}
		
		The diagonalization and quantization of the model was performed in the past using a \emph{Fano diagonalization} method, based on the seminal work of Friedrichs \cite{Friedrichs1937} that was rederived by Fano \cite{Fano1961}. We first review this method.
		
		The initial point is the classical Hamiltonian of the system, where the medium is described as an infinite set of harmonic oscillators interacting with the electromagnetic field through a coupling function $\alpha$. It reads~\cite{Suttorp2004a,Suttorp2004b,Suttorp2007,Philbin2010}:
		\begin{align}
	H&=H_{\text{elm}}+H_{\text{med}}+H_{\text{int}},\label{H_plasm_CC}
	\end{align}
	with
	\begin{subequations}
	\begin{align}
	H_{\text{elm}}&=\int d^3r\left[\frac{1}{2\epsilon_0}\vec\Pi_A^2+\frac{\epsilon_0}{2}\vec A\cdot(c^2\nabla\times\nabla\times\vec A)\right],\\
	H_{\text{med}}&=\int_0^\infty d\nu \int_V d^3r\left[\frac{1}{2}\vec\Pi_X^2+\frac{1}{2}\nu^2\vec X^2\right],\\
	H_{\text{int}}&=\frac{1}{\epsilon_0}\int d^3 r~\vec\Pi_A\cdot\int_0^\infty d\nu~\alpha(\nu,\vec r)\vec X\nonumber\\
	&+\frac{1}{2\epsilon_0}\int_Vd^3 r\left[\int_0^\infty d\nu~\alpha(\nu,\vec r)\vec X\right]^2,
\end{align}
\end{subequations}
where $\vec A$ is the vector potential, $\vec \Pi_A$ is its conjugate momentum,
\begin{align}
	\vec\Pi_A=-\epsilon_0\vec E-\int d\nu~\alpha(\nu,\vec r)\vec X(\nu,\vec r),\label{Eq-PIA}
\end{align}
and $\vec X$,$\vec \Pi_X$ are the canonical variables of the oscillators of the medium. We denote by $\nu$ the frequency of the oscillators, and $V$ is the volume of the medium. The integrals will be written with no boundary when the integration is performed over all space or over all (positive) frequencies.

We consider only an electric interaction but the model can be extended to magnetic interactions by adding a second set of harmonic oscillators \cite{Philbin2010}. The Hamiltonian system described by Eq.~\eqref{H_plasm_CC} was shown \cite{Philbin2010} to imply the macroscopic Maxwell equations, provided that the coupling constant function $\alpha$ is chosen such that
\begin{align}
	\alpha^2(\nu,\vec r)=\frac{2\epsilon_0}{\pi}\nu\epsilon_i(\nu,\vec r),
\end{align}
where $\epsilon_i$ is the imaginary part of the dielectric coefficient of the medium.

In \cite{Suttorp2004a,Suttorp2004b,Suttorp2007,Philbin2010} the diagonalization of the Hamiltonian \eqref{H_plasm_CC} is formulated along with the quantization. It consists in finding a family of bosonic operators $\vec{\hat{C}}(\nu,\vec r)$ satisfying the commutation relations
\begin{subequations}
\begin{align}
	[\hat C_j(\nu,\vec r),\hat C_{j'}^\dag(\nu',\vec r\,')]&=\delta_{jj'}\delta(\vec r-\vec r\,')\delta(\nu-\nu'),\\
	[\hat C_j(\nu,\vec r),\hat C_{j'}(\nu',\vec r\,')]&=0,
\end{align}\label{Eq-commutCC}%
\end{subequations}
such that the Hamiltonian, once quantized, is equal to
\begin{align}
	\hat H=\int d^3r\int d\nu~\hbar\nu~\vec{\hat{C}}^\dag(\nu,\vec r)\cdot\vec{\hat{C}}(\nu,\vec r),\label{HCC}
\end{align}
plus an (infinite) constant that can be dropped.

The construction proceeds by writing the bosonic operator as a linear combination of the canonical variables:
\begin{align}
	&\vec{\hat{C}}(\nu,\vec r)=-\frac{i}{\hbar}\int\! d^3r'\bigg\{\vec{\hat{A}}\cdot \ff^*_{\Pi_A}(\vec r\,'\!,\vec r,\nu)-\vec{\hat{\Pi}}_A\cdot \ff^*_A(\vec r\,'\!,\vec r,\nu)\nonumber\\
	&\!\!+\!\!\int\!d\nu'\big[\vec{\hat{X}}\cdot \ff^*_{\Pi_X}(\vec r\,'\!,\vec r,\nu',\nu)-\vec{\hat{\Pi}}_{X}\cdot \ff^*_X(\vec r\,'\!,\vec r,\nu',\nu)\big]\!\bigg\},\label{Eq-C_expansion}
\end{align}
with some tensors $\ff$. The ansatz \eqref{HCC} implies that the bosonic operators must satisfy:
\begin{align}
\big[\vec{\hat{C}}(\nu,\vec r),\hat H\big]=\hbar\nu~\vec{\hat{C}}(\nu,\vec r).\label{Eq-Heisenberg}
\end{align}
We remark that this is only a necessary condition. In order to make it a sufficient condition it must be complemented with the commutation relations \eqref{Eq-commutCC}. This ensures that the diagonalization is performed canonically, which also allows the quantization to be formally performed before the diagonalization.

Inserting \eqref{H_plasm_CC} and \eqref{Eq-C_expansion} into \eqref{Eq-Heisenberg}, one obtains a system of linear integro-differential equations for the coefficients $\ff$. After some suitable algebraic operations, the general solution of this system can be expressed in terms of a Green tensor and some free undetermined functions. The latter are determined by imposing the commutation relations \eqref{Eq-commutCC}. The solution is still not unique, since there is always the freedom to perform a unitary transformation within the degeneracy subspace. For the remaining free functions one can make a choice that leads to the possibly simplest formulas.

The validity of this construction depends critically on the choice of the ansatz \eqref{HCC}. In order to check whether this ansatz is justified we can consider two different ways to proceed:
\begin{enumerate}
\item One can check whether one recovers the initial Hamiltonian \eqref{H_plasm_CC} when inserting the obtained coefficients $\ff$ into the expression \eqref{Eq-C_expansion} and then into \eqref{HCC}. This check involves relatively complicated calculations and, to our knowledge, it has not been provided in the literature.
\item Another simpler check consists in verifying whether if one takes the limit of zero coupling $\alpha\rightarrow 0$ (i.e., $\epsilon\rightarrow 1$) one obtains the expressions corresponding to the uncoupled medium and the free electromagnetic field. We will show that for a finite medium, these limits lead to the expressions for the uncoupled medium, but one does not recover the electromagnetic field. The conclusion is that the ansatz \eqref{HCC} for the diagonalized Hamiltonian is not valid for a finite medium.
\end{enumerate}

\section{The problem of the no-coupling limit}\label{Sec-problem}

		We analyze specifically the equations in Ref.~\cite{Philbin2010}, Sections~3 and 4 (without the magnetic part of the model). Our discussion can be formulated similarly for the models used in Refs.~\cite{Suttorp2004a,Suttorp2004b,Suttorp2007}, since the quantization procedure followed and the final results are essentially the same as what was described in \cite{Philbin2010} and in the section above. What we present in this Section is based on the assumption of a finite medium. The case of an infinite bulk medium will be discussed specifically in Section~\ref{Sec-infinite-medium}.
		
		In order to check whether the no-coupling limits ($\alpha\rightarrow 0$) of the results presented above are consistent, we need to compare them with the uncoupled model, i.e., when $\hat H_{\text{int}}=0$ in Eq.~\eqref{H_plasm_CC} and $\alpha=0$ in Eq.~\eqref{Eq-PIA}. We expect to find two families of bosonic operators: one corresponding to the free electromagnetic field, given by a linear combination of $\vec A$ and $\vec \Pi_A$, and one corresponding to free matter, given by a linear combination of $\vec X$ and $\vec \Pi_X$. The diagonalization of this uncoupled model indeed yields
		\begin{subequations}
		\begin{align}
			\hat H^0_{\text{elm}}&=\int d^3k\sum_\sigma~\hbar\omega~\hat D^\dag_0(\vec k,\sigma) \hat D_0(\vec k,\sigma)\\
			\hat H^0_{\text{med}}&=\int \!d\nu\!\int_V \!d^3r~\hbar\nu~\vec{\hat{C}}^\dag_0(\nu,\vec r)\cdot\vec{\hat{C}}_0(\nu,\vec r),
		\end{align}\label{Eq-Hdiag0}%
		\end{subequations}
		with $\vec k$ the wave vector in vacuum and $\sigma$ the index of polarization. Here $\hat D_0$ is the operator for photons in free space and $\vec{\hat{C}}_0$ is the operator for the elementary excitations of the free medium,
		\begin{align}
			\vec{\hat{C}}_0=\frac{1}{\sqrt{2\hbar}}\left[\nu^{1/2}\vec{\hat{X}}(\nu,\vec r)+i\nu^{-1/2}\vec{\hat{\Pi}}_{X}(\nu,\vec r)\right].\label{C0}
		\end{align}
		Since the electromagnetic field is not coupled to matter, the electric field operator is the one in vacuum:
		\begin{align}
				\vec{\hat{E}}_0(\vec x)= i\!\!\int\! d^3k\sum_\sigma\sqrt{\frac{\hbar\omega}{2\epsilon_0}}\big[\vec\varphi_{\vec k,\sigma}(\vec x)\hat D_0(\vec k,\sigma)-h.c.\big],\label{Eq-E0}
			\end{align}
			where $\vec\varphi_{\vec k,\sigma}$ are the transverse eigenfunctions of the Laplacian.

	\subsection{Limit for the bosonic operators}
	
	We first analyze the limit of no coupling in the expression of the bosonic operators $\vec{\hat{C}}$ given by the linear combination \eqref{Eq-C_expansion}. In order to check whether they split into the two families of operators $\hat{D}_0$ and $\vec{\hat{C}}_0$ in the no-coupling limit, we need to calculate the coefficients $\ff$. They are given by Eqs.~(80)--(83) of \cite{Philbin2010}:
	\begin{subequations}
		\begin{align}
			\ff_A&=-\frac{i}{\nu}\big[\ff_E\big]_\perp,\\
			\ff_{\Pi_A}&=-\epsilon_0\epsilon\ff_E-\alpha\bar{\bar{h}}_X,\\
			\ff_X&=\frac{i}{\nu}\ff_{\Pi_X},\\
			\ff_{\Pi_X}&=\frac{i\alpha}{\nu}\bigg[1-\bigg(\frac{\nu'}{\nu'-\nu-i0^+}+\frac{\nu'}{\nu'+\nu-i0^+}\bigg)\bigg]\ff_E\nonumber\\
			&-i\nu\bar{\bar{h}}_X\delta(\nu-\nu'),
		\end{align}\label{Eq-coeff}%
		\end{subequations}
		where $\frac{1}{x-i0^+}$ denotes the limit $\lim_{\varepsilon\rightarrow 0^+}\frac{1}{x-i\varepsilon}$. The coefficients are expressed in terms of the two tensors
		\begin{align}
			\ff_E(\vec r\,',\vec r,\nu)&=\sqrt{\frac{\hbar}{2\nu}}\mu_0\nu^2\alpha(\nu,\vec r)\bar{\bar{G}}(\vec r\,',\vec r,\nu),\label{Eq-fe}\\
			\bar{\bar{h}}_X(\vec r\,',\vec r,\nu)&=\sqrt{\frac{\hbar}{2\nu}}\delta(\vec r-\vec r\,')\bar{\bar{\mathds{1}}},
		\end{align}
		where $\bar{\bar{G}}$ is the Green tensor verifying
		\begin{align}
			\left[\nabla\times\nabla\times-\epsilon(\nu,\vec r)\frac{\nu^2}{c^2}\right]\bar{\bar{G}}(\vec r,\vec r\,',\nu)=\delta(\vec r-\vec r\,')\bar{\bar{\mathds{1}}}.\label{Eq-green_EQ}
		\end{align}
		An inspection of these expressions allows one to show that for a finite medium, the no-coupling limit $\alpha\rightarrow 0$ is well defined, and one can calculate it explicitly. We start with Eq.~\eqref{Eq-fe}. The limit of the Green tensor is
		\begin{align}
			\lim_{\alpha\rightarrow 0}\bar{\bar{G}}(\vec r\,',\vec r\,'',\nu)=\bar{\bar{G}}_0(\vec r\,',\vec r\,'',\nu),\label{lim_G}
		\end{align}
		and thus
		\begin{align}
			\lim_{\alpha\rightarrow 0}\ff_E(\vec r\,',\vec r,\nu)=0.
		\end{align}
		Inserting this result in Eqs.~\eqref{Eq-coeff}, we obtain
		\begin{subequations}
		\begin{align}
			\lim_{\alpha\rightarrow 0}\ff_{\Pi_A}&=0,\\
			\lim_{\alpha\rightarrow 0}\ff_A&=0,\\
			\lim_{\alpha\rightarrow 0}\ff_{\Pi_X}&=-i\nu\bar{\bar{h}}_X\delta(\nu-\nu')\nonumber\\
			&=-i\sqrt{\frac{\hbar\nu}{2}}\delta(\vec r-\vec r\,')\delta(\nu-\nu')\bar{\bar{\mathds{1}}},\\
			\lim_{\alpha\rightarrow 0}\ff_X&=\bar{\bar{h}}_X\delta(\nu-\nu')\nonumber\\
			&=\sqrt{\frac{\hbar}{2\nu}}\delta(\vec r-\vec r\,')\delta(\nu-\nu')\bar{\bar{\mathds{1}}}.
		\end{align}
		\end{subequations}
		We can now take the limit in the expression of the bosonic operator \eqref{Eq-C_expansion} and we obtain the expression \eqref{C0}, i.e.,
		\begin{align}
			\lim_{\alpha\rightarrow 0}\vec{\hat{C}}=\vec{\hat{C}}_0.\label{Eq-limit-C}
		\end{align}
		Thus, in the uncoupled limit, the bosonic operator $\vec{\hat{C}}$ becomes the bosonic operator of the free medium $\vec{\hat{C}}_0$ and it contains no information on the bosonic operator of the free electromagnetic field.
	
	\subsection{Limit for the Hamiltonian}
	
	The limit \eqref{Eq-limit-C} can be taken in the diagonal Hamiltonian \eqref{HCC} and compared with Eq.~\eqref{Eq-Hdiag0}. One obtains
	\begin{align}
		\lim_{\alpha\rightarrow 0}\hat{H}&=\!\int\! d^3r\!\int\! d\nu~\hbar\nu~\vec{\hat{C}}_0^\dag(\nu,\vec r)\cdot \vec{\hat{C}}_0(\nu,\vec r)=\hat{H}^0_{\text{med}},
	\end{align}
	which corresponds to the Hamiltonian of the uncoupled medium only; the Hamiltonian of the free electromagnetic field is missing. As mentioned in \cite{Dorier2019a}, this could be expected from the structure of the ansatz \eqref{HCC} which integrates only over the degrees of freedom of matter $(\nu,\vec r)$.
	
	\subsection{Limit for the electric field operator}\label{Sec-E}
	
	In the construction of the diagonalization procedure from \cite{Suttorp2004a,Suttorp2004b,Suttorp2007,Philbin2010} (as well as in the phenomenological approach \cite{Welsch1995,Welsch1996,Welsch1998a,Welsch1998b}), the electric field operator reads
\begin{align}
			\vec{\hat{E}}(\vec x)=\sqrt{\frac{\hbar\mu_0}{\pi c^2}}\int d\nu\int d^3r~\nu^2\epsilon_i^{1/2}(\nu,\vec r)&\bar{\bar{G}}(\vec r,\vec x,\nu)\vec{\hat{C}}(\nu,\vec r)\nonumber\\
			&+h.c.\label{Eq-EwithG}
		\end{align}
		This formula has been widely used in the literature \cite{Matloob1995,Buhmann2003,Suttorp2004a,Suttorp2004b,Suttorp2007,Novotny-Hecht,Fermani2006,Greffet1999,Greffet2000a,Greffet2000b,Greffet2003,Greffet2005,
Greffet2009,Garcia2011,Garcia2014a,Garcia2014b,Hughes2015,Carminati2015,Karanikolas2016,Varguet2016,
Yang2017,Thanopulos2017,Matloob1996,Welsch2001,Dzsotjan2010,Garcia2013,Grimsmo2013,
Zubairy2014,Sinha2014,Rousseaux2016,Castellini2018,Philbin2011,Philbin2014,Philbin2016,Buhmann2015,Yang2019}. In order to show that it cannot be applied to a finite medium, we follow the same procedure: the limits of the Green tensor \eqref{lim_G} and of the bosonic operator \eqref{Eq-limit-C} are regular and well defined. Since the limit $\alpha\rightarrow 0$ is equivalent to $\epsilon_i\rightarrow 0$, the limit yields
		\begin{align}
			\lim_{\alpha\rightarrow 0}\vec{\hat{E}}=0,
		\end{align}
		i.e., the electric field observable would disappear in the uncoupled limit, which of course is not consistent. This conclusion was already presented by \cite{DiStefano2001,Drezet2017a}.
	
	\subsection{Remarks on an infinite bulk medium}\label{Sec-infinite-medium}
	
	The situation for an infinite bulk medium might be different. The uncoupled limit $\alpha\rightarrow 0$ is highly singular, since the dissipation disappears, and it does not seem likely to us that it allows one to recover the expressions of the uncoupled medium \emph{and} electromagnetic field. The singularity of the limit entails that if one does not recover the expressions for the uncoupled fields, it does not mean that formulas for the infinite bulk medium are not correct. At this point we do not make any definite affirmation about the infinite bulk case. In particular, the diagonalization procedure presented in \cite{Dorier2019a} is not applicable to this scenario, since it relies on a M\o ller wave operator which may not exist in an infinite medium. In that case, the spectral structure of the coupled model can be different from the one of the uncoupled model, leading to a possible loss of degrees of freedom. The singular nature of an infinite medium was also evoqued in \cite{Drezet2017c} as a reason for the lack of electromagnetic degrees of freedom in the final results of \cite{Welsch1995,Welsch1996} and subsequent works.
	
	We emphasize that for the applications with nano-structures, the relevant models involve a finite medium. A bulk model can be a good approximation for some specific experimental setups, e.g., if one considers an emitter embedded in the interior of the medium. It is however not appropriate when the emitters are outside the medium, which is a far more common situation, in particular for metallic media.

\section{Justifications and proposed corrections in literature}\label{Sec-justification}

		It has been remarked in several instances in the literature \cite{DiStefano2001,Buhmann2012,Drezet2017a} that the expression of the electric field observable \eqref{Eq-EwithG} cannot yield the free electric field observable in the limit $\epsilon_i\rightarrow 0$. Since this formula is of fundamental importance to study plasmonic structures, some authors have tried to either justify its validity nonetheless, or to correct it.

	\subsection{Addition of a fictitious infinite homogeneous dissipative medium}
	
	An approach to justify the validity of Eq.~\eqref{Eq-EwithG} in the case of a finite medium can be found, e.g., in \cite{Welsch1998b,Buhmann2012,Philbin2016}. It consists of adding artificially to the dielectric coefficient $\epsilon(\nu,\vec r)$ a small homogeneous dissipative background term $\epsilon_\infty(\nu)$, which is set to zero at the very end of the calculations, when one has obtained an expression for a quantity of physical interest, like the spontaneous decay rate of an emitter or the expectation value of a Casimir force. This is a mathematical trick that would allow one to use for practical calculations for a finite medium the expressions obtained for an infinite medium. The difficulty is that the validity of this trick is not easy to justify for the following reasons:
	\begin{itemize}
	\item Certainly, the infinite background medium does not correspond to the considered physical situation. It can only be seen as a mathematical trick, and one has to determine in which sense it can be justified, which does not seem to be an easy task. The procedure clearly shows that taking the limit $\epsilon_\infty(\nu)\rightarrow 1$ at the beginning of the calculation does not give the same result as taking the limit at the end. Thus the exchange of the limit and the intermediate calculations do not commute. Cases like this are difficult to handle mathematically and one has to figure out why one order of the operations can be declared correct and not the other one.
	\item One can ask how to determine whether the results obtained with this trick are correct. In order to make this verification one needs to have an independent method of calculating the desired physical quantities that is known to yield the correct results. The exact diagonalization described in Ref.~\cite{Dorier2019a} provides such an independent method. Until such a comparison is made we cannot make any definite claim on the status of this trick.
	\end{itemize}
	
	\subsection{Addition of a modified free field}\label{Sec_drezet}
	
		Some authors \cite{DiStefano2001,Drezet2017a} have stated that the formula \eqref{Eq-EwithG} is not complete when a finite medium is considered. Although these authors did not derive a formula from the diagonalization of the initial model, they proposed to correct it by adding to the electric field observable \eqref{Eq-EwithG} a contribution $\vec{\hat{E}}_0$,
		\begin{align}
			\vec{\hat{E}}(\vec x)=\vec{\hat{E}}_0(\vec x)+\vec{\hat{E}}_G(\vec x),\label{Eq-EwithG_E0}
		\end{align}
		with $\vec{\hat{E}}_G$ given by Eq.~\eqref{Eq-EwithG}, and $\vec{\hat{E}}_0$ is related to the vacuum field and converges to it in the no-coupling limit. The expression \eqref{Eq-EwithG_E0} has by construction the correct limit when $\epsilon_i\rightarrow 0$. However, there is no justification for claiming that this is the expression that one would obtain from the exact diagonalization of the Hamiltonian \eqref{H_plasm_CC}. In fact, it was shown in \cite{Dorier2019a} that the exact diagonalization in a finite medium leads to an expression for the electric field observable that is different from \eqref{Eq-EwithG_E0} (although it takes a similar form in first order perturbation theory).
		
			\subsection{Exact diagonalization}
			
			Ref.~\cite{Dorier2019a} provides an exact diagonalization of the Hamiltonian in the case of a finite medium. The procedure does not rely on any ansatz of the form \eqref{HCC}. It is based on Lippmann-Schwinger equations to calculate the eigenmodes of the Hamiltonian. These equations can be constructed from the definition of a M\o ller wave operator. The key remark from this procedure is that it ensures the preservation of the uncoupled spectral structure in the diagonalization of the coupled model. Consequently, all the degrees of freedom of the uncoupled model are preserved when the coupling is on.
			
			The diagonal Hamiltonian reads
			\begin{align}
			\hat H\!=\!\int \!d^3k\sum_\sigma~\hbar\omega\hat D^\dag_{\vec k,\sigma} \hat D_{\vec k,\sigma}+\int \!d\nu\!\int_V \!d^3r~\hbar\nu\vec{\hat{C}}^\dag_{\nu,\vec r}\cdot\vec{\hat{C}}_{\nu,\vec r},\label{Eq-Hdiag}
		\end{align}
		with $\hat D$ and $\vec{\hat{C}}$ two families of commuting bosonic operators which tend to the uncoupled ones in the no-coupling limit. The electric field in the exterior of the medium has an expression (Eq.~(25) in \cite{Dorier2019a}) in terms of both families of operators, and it is easily shown that it gives the correct limit as well.
		
		\section{Spontaneous emission and Green identity}\label{Sec-green}
		
		The discussion of this Section is focused on spontaneous emission. Similar arguments apply to the calculation of commutators between the field observables and of correlation functions of the fields, since they use the same Green identity.
		
		The formula \eqref{Eq-EwithG} for the electric field operator has been used in the literature to derive the spontaneous decay rate of the emitter in a plasmonic environment. The conclusion of those calculations is that the decay rate can be expressed by the imaginary part of a Green tensor at the position of the emitter $\vec x_0$:
		\begin{align}
			\Gamma_G(\vec x_0,\omega_0)=\frac{2\omega_0^2}{\hbar\epsilon_0c^2}\left\{\vec d\cdot \text{Im}\left[\bar{\bar{G}}(\omega_0,\vec x_0,\vec x_0)\right]\cdot\vec d\right\},\label{Gamma-lit}
		\end{align}
		with $\omega_0$ the transition frequency of the emitter, $\vec d$ its dipole moment and $\bar{\bar{G}}$ the Green tensor satisfying
		\begin{align}
			\big[\nabla\times\nabla\times-\frac{\nu^2}{c^2}\epsilon(\nu,\vec r)\big]\bar{\bar{G}}(\nu,\vec r,\vec x)=\mathds{1}\delta(\vec r-\vec x).\label{Eq-green1}
		\end{align}
		The decay rate \eqref{Gamma-lit} does not contain inconsistencies when the limit of no coupling is taken. Indeed, in this limit $\bar{\bar{G}}$ tends to the Green tensor in vacuum $\bar{\bar{G}}_0$ satisfying Eq.~\eqref{Eq-green1} with $\epsilon=1$, and one can show \cite{Novotny-Hecht} that we have
		\begin{align}
			\Gamma_0(\vec x_0,\omega_0)=\frac{\omega_0^3|\vec d|^2}{3\pi\hbar\epsilon_0 c^3}.\label{Gamma_vac}
		\end{align}
		The expression \eqref{Gamma-lit} for the decay rate is obtained by inserting the expression \eqref{Eq-EwithG} in the Fermi golden rule \cite{Buhmann2003}. This calculation relies on the validity of Eq.~\eqref{Eq-EwithG}, which is not justified for a finite medium, as shown in Ref.~\cite{Dorier2019a}. Furthermore, we show in the present Section that the usual derivation assuming Eq.~\eqref{Eq-EwithG} also contains discrepancies which have failed to be fully addressed in past works.
		
		\subsection{Fermi golden rule assuming Eq.~\eqref{Eq-EwithG}}
		
		We will now describe how the spontaneous decay rate \eqref{Gamma-lit} is constructed in the literature using the Fermi golden rule and assuming the formula \eqref{Eq-EwithG} for the electric field observable. Note that the same calculation can be performed using the Weisskopf-Wigner theory. We first introduce a coupling operator $\hat W$ between the emitter and the electric field. The Fermi golden rule yields
		\begin{align}
			\Gamma(\vec x_0,\omega_0)=\frac{2\pi}{\hbar^2}\sum_f\left|\mean{f}{\hat W}{i}\right|^2\delta(\omega_f-\omega_0),\label{Gamma-general}
		\end{align}
		where $\ket i$ is the initial state of the system, with the emitter in its excited state $\ket e$ and the plasmonic field in its ground state $\ket \varnothing$. The final states accessible in the resonant approximation correspond to the emitter in its ground state $\ket g$ and one plasmon excited at a frequency $\nu$ and position $\vec r$. Thus,
		\begin{align}
			\ket i=\ket{\varnothing\otimes e},\qquad \ket f=\ket{1_{\nu,\vec r}\otimes g},
		\end{align}		
		and the decay rate \eqref{Gamma-general} is written
		\begin{align}
			\Gamma=\frac{2\pi}{\hbar^2}\!\!\int\! d\nu\!\!\int_V\!\! d^3 r\big|\big\langle 1_{\nu,\vec r}\otimes g\big|\hat W\big|\varnothing\otimes e\big\rangle\big|^2\delta(\nu-\omega_0).\label{Eq-Gamma-sumf}
		\end{align}
		We consider a dipolar interaction $\hat W=-\vec{\hat{d}}\cdot\vec{\hat{E}}$, with $\vec{\hat{d}}=\vec d[\hat\sigma_++\hat\sigma_-]$ and $\hat\sigma_\pm$ the ladder operators for the emitter:
		\begin{align}
			\hat\sigma_+\ket g=\ket e,\quad \hat\sigma_-\ket e=\ket g,\quad\hat\sigma_-\ket g=0.
		\end{align}
		We insert the electric field \eqref{Eq-EwithG} into the operator $\hat W$ and then into the Fermi golden rule \eqref{Eq-Gamma-sumf}. After a few manipulations in the rotating wave approximation, one obtains
		\begin{align}
			\Gamma(\vec x_0,\omega_0)=\frac{2\omega_0^4}{\hbar\epsilon_0 c^4}\left\{\vec d\cdot \bar{\bar{I}}(\vec x_0,\omega_0)\cdot\vec d\right\},
		\end{align}
		where we have used $\mu_0\epsilon_0c^2=1$, and with
		\begin{align}
		\bar{\bar{I}}(\vec x,\omega)=\int_V d^3r~\epsilon_i(\omega,\vec r)\bar{\bar{G}}^T(\omega,\vec r,\vec x)\bar{\bar{G}}^*(\omega,\vec r,\vec x).
		\end{align}
		It is usually stated \cite{Welsch1996,Welsch1998a,Welsch2000,Welsch2007,Buhmann2003,Matloob1996,Welsch2001,
Dzsotjan2010,Garcia2013,Grimsmo2013,Zubairy2014,Sinha2014,Rousseaux2016,Castellini2018,Philbin2014} that this term can be simplified using
		\begin{align}
			\frac{\omega^2}{c^2}\int_V d^3r~\epsilon_i(\vec r)\bar{\bar{G}}^T(\vec r,\vec r_A)\bar{\bar{G}}^*(\vec r,\vec r_B)=\text{Im}~\bar{\bar{G}}(\vec r_A,\vec r_B),\label{Green_identity}
		\end{align}		
		where the frequency dependence is implicit in $\epsilon$ and $\bar{\bar{G}}$. If this were true, the decay rate would simplify into the expression that is mostly used in practice:
		\begin{align}
			\Gamma_G(\vec x_0,\omega_0)=\frac{2\omega_0^2}{\hbar\epsilon_0 c^2}\left\{\vec d\cdot \text{Im}\left[\bar{\bar{G}}(\omega_0,\vec x_0,\vec x_0)\right]\cdot\vec d\right\}.\label{Gamma_litt}
		\end{align}		
		If one assumes the validity of this expression, the question of how to compute the decay rate of the emitter (and also field correlations and commutators) in a given configuration is narrowed to the question of how to find the Green tensor corresponding to this configuration \cite{Matloob1995,Buhmann2003,Novotny-Hecht,Matloob1996,Welsch2001,Dzsotjan2010,Garcia2013,Grimsmo2013,Zubairy2014,Sinha2014,
Rousseaux2016,Castellini2018,Garcia2011,Garcia2014a,Garcia2014b,Hughes2015,Carminati2015,
Karanikolas2016,Varguet2016,Yang2017,Thanopulos2017,NovotnyPRL2006,Greffet2010,
Akselrod2014,Koenderink2017,Sinha2018,Varguet2019,Philbin2011,Philbin2014,Philbin2016,Buhmann2015,Yang2019}. Many analytical and numerical methods have been developed for this specific technique.
		
		\subsection{Green identity for a finite medium}\label{Sec-Green}
		
		The construction above can however not be true for a finite medium. An easy way to see it is by taking the limit $\epsilon\rightarrow 1$ in the Green identity \eqref{Green_identity}. The left-hand side goes to zero while the right-hand side is not zero in general. This is already not the case in vacuum, where the Green tensor reads \cite{Tai1994}
		\begin{align}
			\bar{\bar{G}}_0^{ij}(\omega,\vec r_A,\vec r_B)&=\delta_{ij}g_0+\frac{c^2}{\omega^2}\frac{\partial^2g_0}{\partial r_A^i\partial r_B^j},\\
			g_0(\omega,\vec r_A,\vec r_B)&=\frac{e^{i\frac{\omega}{c}|\vec r_A-\vec r_B|}}{4\pi|\vec r_A-\vec r_B|}.
		\end{align}
		The correct evaluation of the left-hand side of \eqref{Green_identity} brings an extra boundary term, and provided that the Green tensor is reciprocal, i.e., $\bar{\bar{G}}(\vec r_B,\vec r_A)=\bar{\bar{G}}^T(\vec r_A,\vec r_B)$, we have instead the identity \cite{Drezet2017c}
		\begin{align}
			\frac{\nu^2}{c^2}\int_V d^3r~\epsilon_i&(\vec r)\bar{\bar{G}}^T(\vec r,\vec r_A)\bar{\bar{G}}^*(\vec r,\vec r_B)\nonumber\\
			&=\text{Im}~\bar{\bar{G}}(\vec r_A,\vec r_B)+\F(\vec r_A,\vec r_B),\label{TRUE_Green_identity}
		\end{align}
		with
		\begin{align}
			\F(\vec r_A,\vec r_B)&=\frac{1}{2i}\big[\bar{\bar{b}}^T(\vec r_B,\vec r_A)-\bar{\bar{b}}^*(\vec r_A,\vec r_B)\big],\\
			\bar{\bar{b}}(\vec r_A,\vec r_B)&=-\!\int_\mathfrak{B}\!ds~\big(\vec n\times \bar{\bar{G}}^*(\vec x,\vec r_B)\big)^T\big(\nabla\times\bar{\bar{G}}(\vec x,\vec r_A)\big),
		\end{align}
		where $\mathfrak{B}$ denotes a boundary surface which contains $\vec r_A$ and $\vec r_B$ in its interior, $\vec n$ is the outer unit normal vector on the surface, and $ds$ is the surface element. This relation is valid for any reciprocal tensor $\bar{\bar{G}}$ satisfying Eq.~\eqref{Eq-green_EQ}.		
		
		We now recover the correct limit in vacuum, in which case the identity \eqref{TRUE_Green_identity} becomes
		\begin{align}
			\F_0(\vec r_A,\vec r_B)=-\text{Im}~\bar{\bar{G}}_0(\vec r_A,\vec r_B).
		\end{align}
		In conclusion, if the usual formula \eqref{Eq-EwithG} were correct for a finite medium, and if we use the correct Green identity \eqref{TRUE_Green_identity} for this case, the decay rate would read
		\begin{align}
			\Gamma=\frac{2\omega_0^2}{\hbar\epsilon_0 c^2}\left\{\vec d\cdot \left[\text{Im}~\bar{\bar{G}}+\F\right]\cdot\vec d\right\},\label{Gamma_corrected}
		\end{align}
		and not \eqref{Gamma_litt} as stated in the literature \cite{Matloob1995,Buhmann2003,Novotny-Hecht,Matloob1996,Welsch2001,Dzsotjan2010,Garcia2013,Grimsmo2013,Zubairy2014,Sinha2014,
Rousseaux2016,Castellini2018,Garcia2011,Garcia2014a,Garcia2014b,Hughes2015,Carminati2015,
Karanikolas2016,Varguet2016,Yang2017,Thanopulos2017,NovotnyPRL2006,Greffet2010,
Akselrod2014,Koenderink2017,Sinha2018,Varguet2019} (some of these references study the power spectrum, which can be directly linked to the decay rate \cite{HuttnerBarnett1992b}). For a medium interacting weakly with the field (i.e., $\epsilon_i$ close to zero), the correction term $\F$ can be estimated in perturbation theory by
		\begin{align}
			\F=-\text{Im}~\G_0+\mathcal{O}(\epsilon_i),
		\end{align}
		and thus one would have
		\begin{align}
			\Gamma=\frac{2\omega_0^2}{\hbar\epsilon_0 c^2}\left\{\vec d\cdot \left[\text{Im}~\bar{\bar{G}}-\text{Im}~\bar{\bar{G}}_0\right]\cdot\vec d\right\}+\mathcal{O}(\epsilon_i).
		\end{align}
		Since we also have $\G=\G_0+\mathcal{O}(\epsilon_i)$, we would conclude that
		\begin{align}
			\Gamma=\mathcal{O}(\epsilon_i),
		\end{align}
		which implies $\lim_{\epsilon_i\rightarrow 0}\Gamma=0$ instead of the vacuum value \eqref{Gamma_vac}.
		
		One could ask whether the addition of the boundary terms in the Green identity has a significant effect on the decay rate in a non-perturbative scenario. In the next section we calculate it in detail in a one-dimensional configuration where the medium is a slab of metal (which is equivalent to a configuration analyzed, e.g., in \cite{Matloob1995,DiStefano2001,Philbin2016}).

		\subsection{Spontaneous emission in a 1D model assuming Eq.~\eqref{Eq-EwithG}}\label{Sec-Purcell1D}
		
		We consider a 1D model where the material (dissipative and dispersive) medium consists of an homogeneous segment extending from $-\ell$ to $\ell$. Its macroscopic electric response is given by the complex dielectric coefficient $\epsilon$ or equivalently by the (also complex) refractive index $n^2=\epsilon$. This situation corresponds to a special case of a 3D model where only wave vectors $\vec k$ normal to a slab are considered, and where the electric field is constrained in one polarization orthogonal to $\vec k$.
		
		For an observing point $x$, we consider the 1D Green function $G$ satisfying
\begin{align}
	\left[-\partial_x^2-\frac{\omega^2}{c^2}\epsilon(x,\nu)\right]G(x,x_S)=\delta(x-x_S),\label{green_1D}
\end{align}
where $x_S$ is the position of the source. We note $[-L,L]$ the boundaries which must include the coordinates of the source, $x_S$, and of the observer, $x$, and can be sent to infinity. We place arbitrarily the source on the right side of the slab ($x_S\in [\ell,L[$) as in Fig.~\ref{Fig-1Dmain2}. The Green function (with Sommerfeld radiation condition) is given in the Appendix.

\begin{figure}[h]
		\centering
		\includegraphics[width=.7\linewidth]{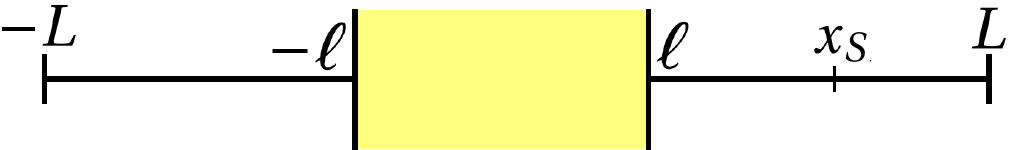}
		\caption{Sketch of the 1D model, with the source placed on the right side of the medium.}\label{Fig-1Dmain2}
		\end{figure}

In 1D, for two possible positions of the source $x_A$ and $x_B$, the Green identity \eqref{TRUE_Green_identity} reads
\begin{align}
	\frac{\omega^2}{c^2}\int dx~\epsilon_i(x)~&G(x,x_A)G^*(x,x_B)\nonumber\\
	&=\text{Im}~G(x_A,x_B)+F(x_A,x_B),\label{Id}
\end{align}
provided that the Green function satisfies the reciprocity condition $G(x,x')=G(x',x)$, which is the case when the Sommerfeld radiation condition is satisfied. Note that the integral in the left-hand side only spans from $-\ell$ to $\ell$ since $\epsilon_i=0$ outside. The boundary term $F$ is
\begin{align}
	F&=\frac{1}{2i}\big[b(x_B,x_A)-b^*(x_A,x_B)\big],\label{F}\\
	b(x_B,x_A)&=-\left[G^*(x,x_B)\partial_x G(x,x_A)\right]_{-L}^L,
\end{align}
and the decay rate reads
\begin{align}
			\Gamma(x_S)=\frac{2\omega_0^2|\vec d|^2}{\hbar\epsilon_0 c^2 S}\left[\text{Im}~G(x_S,x_S)+F(x_S,x_S)\right],\label{Gamma1D}
		\end{align}
		where $S$ is a surface unit introduced for preserving the units of $\Gamma$ in a 1D model. The calculation of the terms in \eqref{Gamma1D} [and in particular the boundary term \eqref{F}] is provided in the Appendix. We obtain
		\begin{align}
			\Gamma=\frac{\omega_0|\vec d|^2}{2\hbar\epsilon_0 c S}\left[1-|A|^2-|D|^2\right],
		\end{align}
		where
\begin{align}
	A&=\frac{4n}{Y}e^{2ikn\ell},\\
	D&=\frac{(n^2-1)}{Y}\big[e^{4ikn\ell}-1\big],\\
	Y&=(n+1)^2-(n-1)^2e^{4ikn\ell}.
\end{align}		
		We remark that $\Gamma$ does not depend on the position of the source in the exterior of the medium, but on the refractive index and on the length of the medium. Consequently, the decay rate of an emitter in the exterior of the slab is constant in space. This is different from the result of the literature \cite{Matloob1995} where the boundary term is missing and the decay rate \eqref{Gamma1D} is reduced to (see the Appendix)
\begin{align}
			\Gamma_G=\frac{\omega_0|\vec d|^2}{\hbar\epsilon_0 cS}\left[1+\text{Re}\left\{De^{-2ik(\ell-x_S)}\right\}\right],\label{Gamma1D_G}
		\end{align}
		which oscillates with the position $x_S$ of the source.
		
Furthermore, we can verify that in the limit $\epsilon\rightarrow 1$ (i.e., $n\rightarrow 1$), we have
\begin{align}
	F(x_A,x_B)\rightarrow -\text{Im}~G_0,\label{Eq-consistency_1D}
\end{align}
where $G_0$ is the 1D Green function satisfying the Green equation in vacuum [Eq.~\eqref{green_1D} with $\epsilon=1$], which implies
\begin{align}
	\Gamma\rightarrow 0.
\end{align}
Hence, Eq.~\eqref{Eq-consistency_1D} is consistent with the identity \eqref{Id}, but the no-coupling limit is not recovered.

In conclusion, this result in a one-dimensional model shows that the missing boundary terms in \eqref{Green_identity} can have a significant effect on the decay rate. Furthermore, the discrepancy with the no-coupling limit implies that one cannot use the expression \eqref{Eq-EwithG} of the electric field to calculate the decay rate (or related quantities) as in Refs.~\cite{Matloob1995,Novotny-Hecht,Buhmann2003,Matloob1996,Welsch2001,Dzsotjan2010,Garcia2013,Grimsmo2013,
Zubairy2014,Sinha2014,Rousseaux2016,Castellini2018,Garcia2011,Garcia2014a,Garcia2014b,
Hughes2015,Carminati2015,Karanikolas2016,Varguet2016,Yang2017,Thanopulos2017,NovotnyPRL2006,
Greffet2010,Akselrod2014,Koenderink2017,Sinha2018,Varguet2019,Philbin2011,Philbin2014,Philbin2016,
Buhmann2015,Yang2019}.

\section{Conclusion}

We have presented a brief summary of the standard approach of the literature to diagonalize and quantize the microscopic plasmon-polariton models. The \emph{Fano diagonalization} technique used relies on an ansatz for the spectral and degeneracy structure of the system which is not adapted when the material medium is of finite size. This conclusion was drawn by studying the limit case where the dissipation in the medium vanishes. We have shown that in this limit, none of the main results of the theory (the Hamiltonian operator, the electric field observable and the creation-annihilation operators) are consistent with the ones of the non-interacting model.

We have also pointed out some inconsistencies in the derivation of the spontaneous emission rate of an emitter in a lossy (but finite) environment when it is done assuming the commonly used expression of the electric field \eqref{Eq-EwithG}. This was shown using the Fermi golden rule, but it can also be applied to the derivation of other quantities such as field correlations and Casimir forces \cite{Greffet2005,Philbin2011,Philbin2014,Philbin2016}.

Ref.~\cite{Dorier2019a} provides a direct (i.e., ansatz-free) diagonalization and quantization of the model with a finite medium, and it differs from the standard results obtained for an infinite medium. We conclude that many phenomena in quantum plasmonics should be reassessed with the revised construction in finite media. In particular, one can try to establish whether some of the formulas constructed for an infinite medium give a good approximation in some regimes.


\subsection*{Acknowledgments}
This work was supported by the French ``Investissements d'Avenir'' program, project ISITE-BFC  I-QUINS (ANR-15-IDEX-03)
and EUR-EIPHI (17-EURE-0002).

\appendix

\section{Green identity in the 1D model}

\begin{figure}[h]
		\centering
		\includegraphics[width=.7\linewidth]{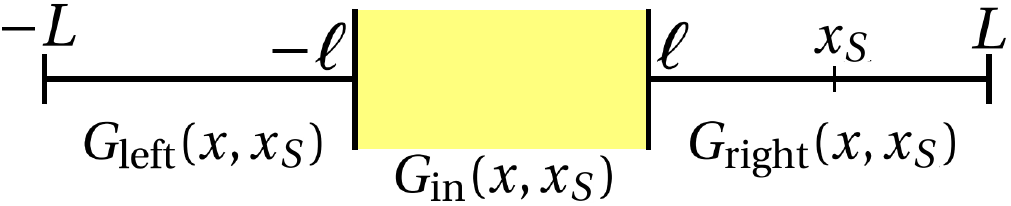}
		\caption{Sketch of the 1D model. The Green function can be split into three contributions, with $x_S$ the position of the source taken on the right side of the medium. The observation point $x$ can be taken either on the left side of the medium, in the medium, or on the right side.}\label{Fig-1Dmain}
		\end{figure}

In this Appendix we perform the calculation of the terms of the Green identity in the 1D model of Section~\ref{Sec-Purcell1D}. Since the observation point $x$ can be placed in three different regions of space, the Green function can be split into three parts. Figure~\ref{Fig-1Dmain} is a sketch of the system with the notation used for each part of the total Green function.
		
We place the source on the right side of the slab ($x_S\in [\ell,L[$). The different parts of the Green function are calculated using the Ohm-Rayleigh expansion and imposing the Sommerfeld radiation condition. It yields
\begin{align}
	G_{\text{left}}(x,x_S)&=\frac{i}{2k}Ae^{-ik(2\ell+x-x_S)},\label{Gleft}\\
	G_{\text{in}}(x,x_S)&=\frac{i}{2k}\big[Be^{-ik(nx-x_S)}+Ce^{ik(nx+x_S)}\big],\\
	G_{\text{right}}(x,x_S)&=\frac{i}{2k}\left[De^{-ik(2\ell-x-x_S)}+e^{ik|x-x_S|}\right],\label{Gright}
\end{align}
where $k=\omega/c$, and
\begin{align}
	A&=\frac{4n}{Y}e^{2ikn\ell},\label{A}\\
	B&=\frac{2(n+1)}{Y}e^{ik(n-1)\ell},\\
	C&=\frac{2(n-1)}{Y}e^{ik(3n-1)\ell},\\
	D&=\frac{(n^2-1)}{Y}\big[e^{4ikn\ell}-1\big],\label{D}
\end{align}
with
\begin{align}
	Y=(n+1)^2-(n-1)^2e^{4ikn\ell}.
\end{align}
For two positions of the source $x_A$ ans $x_B$, we want to calculate the boundary term given by
\begin{align}
	F&=\frac{1}{2i}\big[b(x_B,x_A)-b^*(x_A,x_B)\big],\label{A-F}\\
	b(x_B,x_A)&=-\left[G^*(x,x_B)\partial_x G(x,x_A)\right]_{-L}^L.
\end{align}
Since $L>\ell$, with the notations of Fig.~\ref{Fig-1Dmain} the term $b$ reads
\begin{align}
	b(x_B,x_A)&=G_{\text{left}}^*(x,x_B)\partial_x G_{\text{left}}(x,x_A)\big|_{x=-L}\nonumber\\
	&-G_{\text{right}}^*(x,x_B)\partial_x G_{\text{right}}(x,x_A)\big|_{x=L}.\label{b}
\end{align}
We therefore need two parts of the Green function, one for each position of $x$ close to a boundary. Using the expressions \eqref{Gleft} and \eqref{Gright}, one obtains
\begin{align}
	b=-&\frac{i}{4k}\bigg[(|A|^2+|D|^2)e^{ik(x_A-x_B)}+e^{-ik(x_A-x_B)}\nonumber\\
	&+De^{-ik(2\ell-x_A-x_B)}+D^*e^{ik(2\ell-x_A-x_B)}\bigg]\label{b}
\end{align}
and
\begin{align}
	b^*(x_A,x_B)=-b(x_B,x_A).
\end{align}
We notice that these expressions do not depend on the boundaries $[-L,L]$. Inserting \eqref{b} into \eqref{A-F} gives
\begin{align}
	F(&x_A,x_B)=-\frac{1}{4k}\bigg[(|A|^2+|D|^2)e^{ik(x_A-x_B)}\nonumber\\
	&+e^{-ik(x_A-x_B)}+2\text{Re}\left\{De^{-ik(2\ell-x_A-x_B)}\right\}\bigg].\label{A-F2}
\end{align}
We then choose $x_A=x_B=x_S$,
\begin{align}
	F(x_S,x_S)=-\frac{1}{4k}&\bigg[1+|A|^2+|D|^2\nonumber\\
	&+2\text{Re}\left\{De^{-2ik(\ell-x_S)}\right\}\bigg].\label{Ffinal}
\end{align}
On the other hand, from Eq.~\eqref{Gright} we have the imaginary part of the Green function at the point $x_S$,
\begin{align}
	\text{Im}~G_{\text{right}}(x_S,x_S)=\frac{1}{2k}\left[1+\text{Re}\left\{De^{-2ik(\ell-x_S)}\right\}\right].\label{ImGfinal}
\end{align}
Introducing \eqref{ImGfinal} and \eqref{Ffinal} into the decay rate \eqref{Gamma1D} yields
		\begin{align}
			\Gamma=\frac{\omega_0|\vec d|^2}{2\hbar\epsilon_0 c S}\left[1-|A|^2-|D|^2\right]
		\end{align}
		while \eqref{ImGfinal} gives
		\begin{align}
			\Gamma_G=\frac{\omega_0|\vec d|^2}{\hbar\epsilon_0 cS}\left[1+\text{Re}\left\{De^{-2ik(\ell-x_S)}\right\}\right].
		\end{align}


\end{document}